\documentclass{amsart}

\usepackage{amsmath}
\usepackage{amssymb}
\usepackage{amsfonts}
\usepackage{latexsym}
\usepackage{color}

\newtheorem{theo}{Theorem}[section]
\newtheorem{lem}[theo]{Lemma}
\newtheorem{prop}[theo]{Proposition}
\newtheorem{cor}[theo]{Corollary}
\newtheorem{lemma}[theo]{Lemma}

\newtheorem{definition}[theo]{Definition}
\newtheorem{example}[theo]{Example}
\newtheorem{remark}[theo]{Remark}

\newcommand{\betheo}{\begin{theo}$\!\!\!$ }
\newcommand{\entheo}{\end{theo}}

\newcommand{\becor}{\begin{cor}$\!\!\!$  }
\newcommand{\encor}{\end{cor}}

\newcommand{\belem}{\begin{lem}$\!\!\!${\bf .} }
\newcommand{\enlem}{\end{lem}}

\newcommand{\beprop}{\begin{prop}$\!\!\!${\bf .} }
\newcommand{\enprop}{\end{prop}}

\newcommand{\bedefi}{\begin{definition}$\!\!\!$ \rm }
\newcommand{\findefi}{ \end{definition}}

\newcommand{\beex}{\begin{example}$\!\!\!$ \rm }
\newcommand{\enex}{ \end{example}}

\newcommand{\berem}{\begin{remark}$\!\!\!$ \rm }
\newcommand{\enrem}{ \end{remark}}

\newcommand{\be}{\begin{equation}}
\newcommand{\en}{\end{equation}}

\newcommand{\bea}{\begin{eqnarray}}
\newcommand{\ena}{\end{eqnarray}}

\newcommand{\beano}{\begin{eqnarray*}}
\newcommand{\enano}{\end{eqnarray*}}

\newcommand{\bee}{\begin{enumerate}}
\newcommand{\ene}{\end{enumerate}}

\newcommand{\bei}{\begin{itemize}}
\newcommand{\eni}{\end{itemize}}

\newcommand{\betab}{\begin{tabular}}
\newcommand{\entab}{\end{tabular}}

\newcommand{\bd}{\begin{displaymath}}

\newcommand{\ad}{^{\mbox{\scriptsize $\dag$}}}

\newcommand{\up}{\raisebox{0.7mm}{$\upharpoonright \,$}}%

\def\D{{\mathcal D}}
\def\E{{\mathcal E}}

\def\H{{\mathcal H}}

\def\L{{\mathcal L}}
\def\M{{\mathcal M}}

\def\J{\relax\ifmmode {\mathcal J}\else${\mathcal J}$\fi}
\def\x{\relax\ifmmode {\mbox{*}}\else*\fi}

\def\MM{{\mathfrak M}}
\def\NG{{\mathfrak N}}

\newcommand{\mc}{\mathcal}
\newcommand{\mb}{\mathbb}

\newcommand{\LD}{{\L}\ad(\D)}

\newcommand{\BH}{{\mc B}(\H)}

\newcommand{\w}{{\rm w}}
\newcommand{\cu}{{\rm c}}

\newcommand{\bic}[1]{{#1}''_{\w\cu}}
\newcommand{\wcom}{\MM'_\w\D\subset \D}

\def\dag{\dagger}

\newcommand{\ip}[2]{\left\langle{#1}\right|\left.{#2}\right\rangle}

\def\OL{\relax\ifmmode {\sf L}\else{\textsf L}\fi}
\def\OR{\relax\ifmmode {\sf R}\else{\textsf R}\fi}

\begin{document}
\title{Bicommutants of reduced unbounded operator algebras}

\author[Bagarello]{Fabio Bagarello}
\address
{Dipartimento di Metodi e Modelli Matematici,
 Facolt\`a d'Ingegneria, Universit\`a di Palermo,
I-90128 Palermo, Italy} \email{bagarell@unipa.it}
\author[Inoue]{Atsushi Inoue}
\address{
Department of Applied Mathemathics, Fukuoka University, 814-0180, Japan} \email{a-inoue@fukuoka-u.ac.jp}

\author[Trapani]{Camillo Trapani}
\address{
Dipartimento di Matematica ed Applicazioni , Universit\`a di Palermo, I-90123 Palermo, Italy}
\email{trapani@unipa.it}

\thanks{This work was supported by CORI, Universit\`a di Palermo.}

\begin{abstract}The unbounded bicommutant ${\bic{(\MM_{E'})}}$ of the {\em reduction} of an O*-algebra $\MM$ via a given projection
$E'$ weakly commuting with $\MM$ is studied, with the aim of finding conditions under which the reduction of a
GW*-algebra is a GW*-algebra itself. The obtained results are applied to the problem of the existence of
conditional expectations on O*-algebras.
\end{abstract}

 \maketitle

\section{Introduction and Preliminaries}
If $\MM$ is a von Neumann algebra and $E'$ a projection on the commutant $\MM'$ of $\MM$, then it is well known
that the {\em reduced} algebra $\MM_{E'}=\{XE';\, X \in \MM\}$ is again a von Neumann algebra. The notion of von
Neumann algebra has been generalized to unbounded operator algebras by introducing several classes of
O*-algebras, such as GW*-algebras, EW*-algebras, etc. It is then natural and important for applications to pose
the question if the {\em reduced} algebra of a GW*-algebra via a projection $E'$ picked in its weak bounded
commutant is again a GW*-algebra. In order to answer this question we undertake a more general study of the
reduction process for O*-algebras. We begin with reviewing the basic definitions and properties of O*-algebras
needed in this paper and refer to \cite{powers, schmudgen} for more details.

Let $\H$ be a Hilbert space with inner product $\ip{\cdot}{\cdot}$ and $\D$ a dense subspace of $\H$. We denote
by $\LD$ the set of all linear operators $X$ defined in $\D$ such that $X\D \subset \D$, the domain $D(X^*)$ of
its adjoint contains $\D$ and $X^* \D \subset \D$. Then $\LD$ is a *-algebra with the usual operations: $X+Y$,
$\alpha X$, $XY$ and the involution $X \mapsto X\ad := X^*{\up\D}$. A *-subalgebra of $\LD$ is called an
O*-algebra on $\D$. In this paper we will assume that an O*-algebra always contains the identity $I$.

Let $\MM$ be an O*-algebra on $\D$. The {\em graph topology} $t_\MM$ on $\D$ is the locally convex topology
defined by the family $\{ \|\cdot\|_X;\, X \in \MM\}$ of seminorms: $\|\xi\|_X= \|X\xi\|$, $\xi \in \D$. If the
locally convex space $\D[t_\MM]$ is complete, then $\MM$ is said to be {\em closed}. In particular, if
$\D[t_\MM]$ is a Fr\'echet space, then $\D$ is called the Fr\'echet domain of $\MM$. More in general, we denote
by $\widetilde{\D}(\MM)$ the completion of the locally convex space $\D[t_\MM]$ and put
$$ \widetilde{X}:= \overline{X}\up\widetilde{\D}(\MM) \quad\mbox{ and } \widetilde{\MM}:=\{\widetilde{X}: X \in
\MM\}.$$ Then $\widetilde{\D}(\MM)=\bigcap_{X\in \MM}D(\overline{X})$ and $\widetilde{\MM}$ is a closed
O*-algebra on $\widetilde{\D}(\MM)$ which is called the {\em closure} of $\MM$, since it is the smallest closed
extension of $\MM$. We next recall the notion of {self-adjointness} of $\MM$. If $\D=\D^*(\MM) := \bigcap_{X\in
\MM}D(X^*)$, then $\MM$ is said to be {\em self-adjoint}. If $\widetilde{\D}(\MM)=\D^*(\MM)$, then $\MM$ is said
to be {\em essentially self-adjoint}. It is clear that
\begin{align*}
& \D\subset \widetilde{\D}(\MM) \subset \D^*(\MM)\\
& X\subset \widetilde{X} \subset {X\ad}^*, \quad \forall X \in \MM.
\end{align*}
The {\em weak commutant} $\MM'_\w$ of $\MM$ is defined by
$$\MM'_\w=\{C\in \BH: \ip{CX\xi}{\eta}=\ip{C\xi}{X\ad\eta}, \; \forall X\in \MM,\,\forall \xi, \eta\in \D \},$$
where $\BH$ is the *-algebra of all bounded linear operators on $\H$. Then $\MM'_\w$ is a weak-operator closed
*-invariant subspace of $\BH$ (but it is not, in general, a von Neumann algebra) and $(\widetilde{\MM})'_\w
=\MM'_\w$. If $\MM'_\w\D\subset \D$, as it happens for self-adjoint $\MM$,  then $\MM'_\w$ is a von Neumann
algebra. In this paper we will use the following bicommutants of $\MM$:
$$(\MM'_\w)' =\{A \in \BH: \, AC=CA,\, \forall C \in \MM'_\w\},$$
$$\bic{\MM}:= (\MM'_\w)'_\cu=\{X\in \LD:\, \ip{CX\xi}{\eta}=\ip{C\xi}{X\ad\eta}, \;
\forall C\in \MM'_\w,\,\forall \xi, \eta\in \D \}.$$ Then, $(\MM'_\w)'$ is a von Neumann algebra on $\H$ and
$\bic{\MM}$ is a $\tau_{s^*}$-closed O*-algebra on $\D$ such that $\MM \subset \bic{\MM}$ and $(\bic{\MM})'_\w =
\MM'_\w$, where the strong*-topology $\tau_{s^*}$ is defined by the family $\{p_\xi^*(\cdot); \xi \in \D\}$ of
seminorms:
$$p_\xi^*:=\|X\xi\|+\|X\ad\xi\|,\quad X \in \LD.$$
Furthermore, if $\MM'_\w\D\subset \D$, then
$$\bic{\MM} =\{X \in \LD:\, \overline{X} \mbox{ is affiliated with }(\MM'_\w)'\}=
\overline{(\MM'_\w)'\up\D}^{\tau_{s^*}}\cap \LD. $$

An O*-algebra $\MM$ on $\D$ is called a {\em GW*-algebra} if $\wcom$ and $\MM=\bic{\MM}$.

Let $\MM$ be a closed O*-algebra on $\D$ such that $\wcom$ and let $E'$ be a projection in the von Neumann
algebra $\MM'_\w$. Let $\MM_{E'}$ denote the {\em reduced algebra} $\MM_{E'}= \{XE':\,X\in \MM\}$. Then it is
known \cite{dixmier} that
\begin{align*}
&(\MM_{E'})'_\w= (\MM'_\w)_{E'},\\
&((\MM_{E'})'_\w)'=((\MM'_\w)')_{E'}.
\end{align*}
But the following question is open

\medskip \noindent{\bf Question:} Does the equality $\bic{(\MM_{E'})}=(\bic{\MM})_{E'}$ hold?

\medskip If the answer to this question is affirmative, then the reduced algebra $\MM_{E'}$ of a given
GW*-algebra is again a GW*-algebra. This question will be considered in Section 2, where we will show that if
$\D$ is the Fr\'echet domain of $\MM$ and the linear span $\large\langle  \MM'_\w E'\D  \large\rangle$ of the
$\MM$-invariant subspace $\MM'_\w E'\D$ is essentially self-adjoint, then $\bic{(\MM_{E'})}=(\bic{\MM})_{E'}$. An
$\MM$-invariant subspace $\M$ of $\D$ is called {\em essentially self-adjoint} if the O*-algebra
$\MM\up\M:=\{X\up\M: \, X\in \MM\}$ of $\M$ is essentially selfadjoint.

Furtheremore, we will show that, if the graph topology $t_\MM$ is defined by a sequence $\{T_n\}$ of essentially
self-adjoint operators of $\MM$, whose spectral projections leave the domain $\D$ invariant, then again
$\bic{(\MM_{E'})}=(\bic{\MM})_{E'}$.

In Section 3 we shall apply the results of Section 2 to the study of conditional expectations for O*-algebras.

\section{Main results}

Let $\MM$ be a closed O*-algebra on $\D$ such that $\wcom$ and let  $E'$ be a projection in the von Neumann
algebra $\MM'_\w$. In this Section we look for conditions for the equality $(\bic{\MM})_{E'}=\bic{(\MM_{E'})}$ to
hold. Since $(\MM_{E'})'_\w= (\MM'_\w)_{E'}$, it follows that \begin{equation}\label{2.1}
 (\bic{\MM})_{E'}\subset ((\MM'_\w)_{E'})'_\cu=
\bic{(\MM_{E'})}.\end{equation} Hence we need only to prove the converse inclusion: $\bic{(\MM_{E'})}\subset
(\bic{\MM})_{E'}$. An element $X$ of $\bic{(\MM_{E'})}$ is an operator on $E'\D$, and so to show that $X \in
(\bic{\MM})_{E'}$ we need to extend $X$ to an operator on $\D$. We will consider this extension problem below.

Let $Z$ denote the {\em central support} of $E'$ i.e., $Z$ is the projection onto the closure of the subspace
$\large\langle \MM'_\w E'\H \large\rangle$. We define ${\mc E} := \large\langle \MM'_\w E'\D \large\rangle \oplus
(I-Z)\D$. Then $\large\langle  \MM'_\w E'\D \large\rangle\subset ZD$ and $\E\subset \D$.

For $X\in \bic{(\MM_{E'})}$, we put

$$X_e\left( \sum_k C_kE'\xi_k+(I-Z)\eta\right):= \sum_k C_kXE'\xi_k, \quad C_k \in \MM'_\w, \xi_k, \eta\in
\D.$$ Then we have the following
\begin{lemma} $X_e\in {\L}\ad(\E)$ and $(\bic{(\MM_{E'})})_e := \{X_e: X\in \bic{(\MM_{E'})}\}$ is an
O*-algebra on $\E$.
\end{lemma}
\begin{proof} For any $X \in \bic{(\MM_{E'})}$, $C_k, K_j \in \MM'_\w$ and $\xi_k, \zeta_j, \eta, \eta_1 \in \D$
we have
\begin{eqnarray*}
\lefteqn{\ip{X_e\left( \sum_k C_kE'\xi_k+(I-Z)\eta\right)}{ \sum_j
K_jE'\zeta_j+(I-Z)\eta_1}}\\&=&\sum_{k,j}\ip{C_kXE'\xi_k}{K_jE'\zeta_j}\\
&=& \sum_{k,j}\ip{E'K_j^*C_kE'XE'\xi_k}{E'\zeta_j}\\
&=& \sum_{k,j}\ip{E'K_j^*C_kE'\xi_k}{X\ad E'\zeta_j}\\
&=& \ip{\sum_k C_kE'\xi_k}{\sum_j K_jX\ad E'\zeta_j}\\
&=& \ip{\sum_k C_kE'\xi_k+(I-Z)\eta}{(X_e)\ad\left(\sum_j K_jE'\zeta_j+(I-Z)\eta_1\right)}
\end{eqnarray*}
which implies that $X_e$ is well-defined and $(X_e)\ad= (X\ad)_e$. It is clear that $X_e$ is a linear operator on
$\E$. Hence $X_e \in  {\L}\ad(\E)$. Moreover, as it is easily seen, $(XY)_e = X_e Y_e.$ Hence,
$e(\MM_{E'}):=(\bic{(\MM_{E'})})_e$ is an O*-algebra on $\E$.
\end{proof}

Since $e(\MM_{E'})$ is an O*-algebra on $\E$, its closure is an O*-algebra on $\widetilde{\E}$. It is easily seen
that $\widetilde{\E}=\overline{\large\langle \MM'_\w E'\D \large\rangle}^{t_{e(\MM_{E'})}} \oplus (I-Z)\D$. On
the other hand, $\D=Z\D\oplus (I-Z)\D$; hence, if $\overline{\large\langle \MM'_\w E'\D
\large\rangle}^{t_{e(\MM_{E'})}}=Z\D$, we can extend any $X \in \bic{(\MM_{E'})}$ to $\D$.

\betheo \label{thm_2.2}Let $\MM$ be a closed O*-algebra on $\D$ such that $\wcom$ and let $E'$ be a projection in
the von Neumann algebra $\MM'_\w$. Suppose that $\widetilde{\E}(\bic{(\MM_{E'})})_e)=\D$ or, equivalently,
$\overline{\large\langle \MM'_\w E'\D \large\rangle}^{t_{e(\MM_{E'})}}=Z\D$. Then,
$(\bic{\MM})_{E'}=\bic{(\MM_{E'})}$.\entheo
\begin{proof} By the assumption, for every $X\in \bic{(\MM_{E'})}$, $X_e$ extends to an operator $\widetilde{X_e}$ on $\D$; that is
$\widetilde{X_e}=\overline{X_e}\up\D$ and $(\widetilde{X_e})_{E'}=X$. Furthermore, we have $\widetilde{X_e}\in
\bic{\MM}$. Indeed, for every $C \in \MM'_\w$, we have
\begin{eqnarray*} CX_e\left(\sum_k C_kE'\xi_k+(I-Z)\eta\right)&=&C\sum_k C_kXE'\xi_k\\
&=& \sum_k C C_kXE'\xi_k\\
&=& X_eC\left(\sum_k C_kE'\xi_k+(I-Z)\eta \right),
\end{eqnarray*}
for every $C_k\in \MM'_\w,\, \xi_k, \eta \in \D$. Hence, $CX_e =X_eC$ on $\E$ and so $C\widetilde{X_e}
=\widetilde{X_e}C$. Thus $\widetilde{X_e}\in \bic{\MM}$ and $X=(\widetilde{X_e})_{E'} \in (\bic{\MM})_{E'}$. In
conclusion, $\bic{(\MM_{E'})}\subset (\bic{\MM})_{E'}$ and by \eqref{2.1} the equality follows.
\end{proof}
The next step consists, of course, in looking for situations where the conditions of Theorem \ref{thm_2.2} are
satisfied. We begin with the following
\begin{lemma}\label{lem_2.3} Let $\MM$ and $E'$ be as above. Suppose that
\begin{itemize}
\item[(i)]for every $X \in \bic{(\MM_{E'})}$ there exists an element $Y$ of $\bic{\MM}$ such that
$$\|X_e\xi\|\leq \|Y\xi\|, \quad \forall \xi \in \E$$ (equivalently, $X_e\ad X_e\leq Y\ad Y$ on $\E$)
\item[(ii)] $\overline{\large\langle \MM'_\w E'\D \large\rangle}^{t_{\bic{\MM}}}=Z\D$.
\end{itemize}
Then, $\bic{(\MM_{E'})}=(\bic{\MM})_{E'}$.
\end{lemma}
\begin{proof}The conditions (i) and (ii) immediately imply that
$$Z\D \subset \overline{\large\langle \MM'_\w E'\D
\large\rangle}^{t_{e(\MM_{E'})}}.$$ Hence
$$ \overline{\large\langle \MM'_\w E'\D
\large\rangle}^{t_{e(\MM_{E'})}}=Z\D.$$ By Theorem \ref{thm_2.2}, we get $\bic{(\MM_{E'})}=(\bic{\MM})_{E'}$.
\end{proof}
\berem A comment is in order for condition (ii) of Lemma \ref{lem_2.3}. Let us in fact consider the following
statements:
\begin{itemize}
\item[(i)] The $\bic{\MM}$-invariant subspace $\large\langle \MM'_\w E'\D
\large\rangle$ is essentially self-adjoint for $\bic{\MM}$.
\item[(ii)] $\overline{\large\langle \MM'_\w E'\D \large\rangle}^{t_{\bic{\MM}}}=Z\D$.
\end{itemize}
Then (i) $\Rightarrow$ (ii). In particular if $\MM$ is self-adjoint, then the two conditions are equivalent. For
the proof we refer to \cite{powers}. \enrem

 \betheo \label{thm_2.4}Let $\MM$ be a closed O*-algebra $\D$ with $\D$ a Fr\'echet domain of $\MM$. Assume
that $\MM'_\w\D \subset \D$. Let $E'$ be a projection in $\MM'_\w$ and assume that $\large\langle \MM'_\w E'\D
\large\rangle$ is essentially self-adjoint for $\bic{\MM}$. Then, $\bic{(\MM_{E'})}=(\bic{\MM})_{E'}$. \entheo
\begin{proof}
Since $\D[t_\MM]$ is a Fr\'echet space, the topology $t_\MM$ is defined by a sequence $\{T_n\}$ of elements of
$\MM$ satisfying
$$\|\xi\|\leq \|T_1\xi\| \leq \|T_2\xi\|, \ldots, \quad \forall \xi \in \D.$$
Any $X \in \LD$ is a closed linear operator on the Fr\'echet space $\D[t_\MM]$ into the Hilbert space $\H$, which
implies, by the closed graph theorem, that
 the topology $t_\MM$ is equivalent to the graph topology $t_{\LD}$ defined
by $\LD$. This in turn implies that, for every $X \in \LD$ there exist $n_0 \in {\mb N}$ and $\gamma>0$ such that
\begin{equation}\label{2.2} \|X\xi\|\leq \gamma \|T_{n_0}\xi\|, \quad \forall \xi \in \D.\end{equation} Take now
an arbitrary $X\in \bic{(\MM_{E'})}$. Since $XE'\in \LD$, it follows from \eqref{2.2} that
\begin{equation}\label{2.3}\|XE'\xi\|\leq \gamma \|T_{n_0}\xi\|, \quad \forall \xi \in \D.\end{equation}
Since $\MM'_\w$ is a von Neumann algebra, for every $C\in \MM'_\w$ and $\xi \in \D$, we have
\begin{eqnarray*}\|X_e(CE'\xi)\|^2&=& \|CXE'\xi\|^2\\
&=&\ip{E'C^*CE'XE'\xi}{XE'\xi}\\
&=&\|(E'C^*CE')^{1/2}XE'\xi\|^2\\
&=&\|XE'(E'C^*CE')^{1/2}\xi\|^2\\
&\leq&\gamma^2 \|T_{n_0}(E'C^*CE')^{1/2}\xi\|^2\\
&=&\gamma^2 \|(E'C^*CE')^{1/2}T_{n_0}\xi\|^2\\
&=&\gamma^2\|CE'T_{n_0}\xi\|^2\\
&=&\gamma^2\|T_{n_0}CE'\xi\|^2.
\end{eqnarray*}
Hence, the condition (i) of Lemma \ref{lem_2.3} is satisfied. By the same Lemma we then get the equality
$\bic{(\MM_{E'})}=(\bic{\MM})_{E'}$.
\end{proof}

\begin{lemma} \label{lemma_2.5}Let $\MM$ be a closed O*-algebra on a Fr\'echet domain $\D$ in $\H$ such that $t_\MM$ is defined by a sequence $\{T_n\}$ of
essentially self-adjoint operators of $\MM$ whose spectral projections belong to $\MM$.

Let $\M$ be an $\MM$-invariant subspace of $\D$ such that the projection $P_\M$ onto $\overline{\M}$ belongs to $\MM_w'$.

Then $\M$ is essentially self-adjoint for $\MM$ or, equivalently, $P_\M\D=\overline{\M}^{t_\MM}$.

\end{lemma}

\begin{proof}

First we observe that $\overline{\M}^{t_\MM}\subset P_\M\D$. Hence we only need to show that the converse
inclusion also holds. Take an arbitrary $\xi\in\D$. Then there exists a sequence $\{\xi_n\}$ in $\M$ such that
$\lim_{n\to\infty}\|\xi_n-P_\M\xi\|=0$. By \eqref{2.3}, for every $X \in \MM$, there exists $n_0 \in {\mb N}$ and
$\gamma>0$ such that \begin{equation}\label{2.4a} \|X\xi\|\leq \gamma \|T_{n_0}\xi\|, \quad \forall \xi \in
\D.\end{equation} By the assumption, $\overline{T_{n_0}}$ is self-adjoint. Let $
\overline{T_{n_0}}=\int_{-\infty}^{\infty} \lambda d E_{T_{n_0}}(\lambda)$ be the spectral resolution of
$\overline{T_{n_0}}$.We put $E_k:=E_{T_{n_0}}(k)-E_{T_{n_0}}(-k)$, $k \in {\mb N}$. Then, by assumption $E_k\up
\D\in \MM$, $\forall k \in {\mb N}$ and so, $E_k\xi_n\in \M$, $\forall k,n \in {\mb N}$ and $\lim_{n\to\infty}
\|E_k\xi_n - E_kP_\M\xi\|=0$. Furthermore, by \eqref{2.4a} we have
$$ \|XE_k\xi_n - XE_k\xi_m\|\leq \gamma \|T_{n_0}E_k(\xi_n- \xi_m)\|\to 0, \;\mbox{as }n,m \to \infty.$$
Hence $E_kP_\M\xi \in D(\overline{X\up \M})$.

Furthermore,
\begin{eqnarray*}\|XE_kP_\M\xi - XP_\M\xi\|&\leq& \gamma \|T_{n_0}E_kP_\M\xi - T_{n_0}P_\M\xi\|\\
&=&\gamma \|P_\M(E_kT_{n_0}\xi-T_{n_0}\xi)\|\\
&\leq&\gamma\|E_kT_{n_0}\xi- T_{n_0}\xi\| \to 0 \;\mbox{as }k \to \infty,
\end{eqnarray*}
which implies that $P_\M\xi \in D(\overline{X\up \M})$. Hence,
$$ P_\M\xi  \in \bigcap_{X \in \MM} D(\overline{X\up \M})=\overline{\M}^{t_\MM}.$$
Thus $P_\M\D= \overline{\M}^{t_\MM}$. This is equivalent to $\M$ being essentially self-adjoint for $\MM$, since
$\MM$ is clearly self-adjoint.

\end{proof}

By Theorem \ref{thm_2.4} and Lemma \ref{lemma_2.5} we have the following

\betheo \label{thm_2.5} Let $\MM$ be a closed O*-algebra such that $t_\MM$ is defined by a sequence $\{T_n\}$ of
essentially self-adjoint operators of $\MM$ whose spectral projections leave the domain $\D$ invariant. Let $E'$
be a projection in $\MM'_\w$. Then $\bic{(\MM_{E'})}=(\bic{\MM})_{E'}$. \entheo
\begin{proof}
It can be shown that $\MM$ is a self-adjoint O*-algebra on the Fr\'echet domain $\D$, so
that $\MM_w'\D\subset\D$, $t_{\bic{\MM}}=t_\MM$ and $\bic{\MM}$ is a self-adjoint O*-algebra on the
Fr\'echet domain $\D$. Furthermore
$$
P_{\langle\M_w' E'\D\rangle}=Z\in \MM_w'\cap(\MM_w')' =(\bic{\MM})'_\w\cap((\bic{\MM})'_\w)'.$$ Moreover, since
the spectral projections of the operators $T_n$ leave the domain $\D$ invariant, they automatically belong to
$\bic{\MM}$. Hence, the self-adjoint O*-algebra $\bic{\MM}$ and the $\bic{\MM}$-invariant subspace $\langle\M_w'
E'\D\rangle$ of $\D$ satisfy all the conditions in Lemma \ref{lemma_2.5}, and so $Z\D=\overline{\langle\M_w'
E'\D\rangle}^{t_{\bic{\MM}}}$. Then, by Theorem \ref{thm_2.4}, $\bic{(\MM_{E'})}=(\bic{\MM})_{E'}$.

\end{proof}

\begin{cor}Let $T$ be an essentially self-adjoint operator in $\H$ and  $\MM$ be a self-adjoint O*-algebra on $\D^\infty (\overline{T}):=\bigcap_{n\in {\mb
N}}D(\overline{T}^n)$, containing $T$. Let $E'$ be a projection in $\MM'_\w$. Then
$\bic{(\MM_{E'})}=(\bic{\MM})_{E'}$.
\end{cor}
\begin{proof}  The spectral projections $E_T(\lambda)$, $\lambda \in {\mb R}$, of $\overline{T}$ satisfy:
\begin{itemize}
\item $E_T(\lambda)\in (\MM'_\w)', \quad \forall \lambda \in {\mb R}$;
\item $E_T(\lambda)\H\subset \D^\infty (\overline{T}) , \quad \forall \lambda \in {\mb R}$.
\end{itemize}
The statement then follows from Theorem \ref{thm_2.5}.
\end{proof}

Apart from GW*-algebras,  another unbounded generalization of von Neumann algebras is provided by the notion of
{\em extended} W*-algebras, shortly EW*-algebras, defined as follows: A closed O*-algebra $\MM$ on $\D$ is said
to be an {\em EW*-algebra} if $(I+X\ad X)^{-1}$ exists in $\MM_{\rm b}:= \{A\in \MM:\, \overline{A} \in \BH\}$,
for every $X\in \MM$ and $\overline{\MM_{\rm b}}:= \{\overline{A}\in \MM:\,  A \in \MM_{\rm b}\}$ is a von
Neumann algebra \cite{inoue1}. It is easily shown that every EW*-algebra on a Fr\'echet domain satisfies the
conditions of Theorem \ref{thm_2.5} (remind that every symmetric element of an EW*-algebra is essentially
self-adjoint). Hence we have the following
\begin{cor} Let $\MM$ be an EW*-algebra on the Fr\'echet domain $\D$ in Hilbert space $\H$ and
$E'$ a projection in $\MM'_\w$. Then, $\bic{(\MM_{E'})}=(\bic{\MM})_{E'}$.
\end{cor}
\newpage
\section{Applications}

In this section we show how to use the results of Section 2 in the analysis of the existence of conditional
expectations for O*-algebras, which were first studied in \cite{inoue3, inoue4}.

Let $\MM$ be a given O*-algebra on $\D$ in $\H$ with a strongly cyclic vector $\xi_0$. Here $\xi_0\in\D$ is said
to be strongly cyclic for $\MM$ if $\overline{\MM\xi_0}^{t_\MM}=\D$. With no loss of generality we will assume
that $\|\xi_0\|=1$. Let $\NG$ be a O*-subalgebra of $\MM$. A map $\E$ of $\MM$ onto $\NG$ is said to be a
conditional expectation of $(\MM,\xi_0)$ w.r.t. $\NG$ if it satisfies the following conditions:
\begin{itemize}

\item[(i)] $\E(X)^\dagger=\E(X^\dagger)$,  $\forall\,X\in\MM$,  and $\E(A)=A$, $\forall A\in\NG$;

\item[(ii)] $\E(XA)=\E(X)A$ and $\E(AX)=A\E(X)$,  $\forall\,X\in\MM$, $\forall\,A\in\NG$;

\item[(iii)] $\omega_{\xi_0}(\E(X))=\omega_{\xi_0}(X)$, $\forall\,X\in\MM$, where $\omega_{\xi_0}$ is a state on $\MM$ defined as $\omega_{\xi_0}(X)=\ip{X\xi_0}{\xi_0}$, $X\in\MM$.

\end{itemize}

In the case of von Neumann algebras Takesaki \cite{take} characterized the existence of conditional expectations
using Tomita's modular theory. Thus a conditional expectation does not necessarily exist for a general von
Neumann algebra. Hence, in \cite{inoue3, inoue4} Ogi, Takakura and one of us considered  a linear map $\E$ of a
$\dagger$-invariant subspace $D(\E)$ of $\MM$ onto $\NG$ satisfying the above conditions (i), (ii) and (iii) on
$D(\E)$. Such a map is called an {\em unbounded conditional expectation} of $(\MM,\xi_0)$ w.r.t. $\NG$, and it
was shown that there exists the largest unbounded conditional expectation $\E_\NG$ of $(\MM,\xi_0)$ w.r.t. $\NG$.
Furthermore,  the existence of conditional expectation was characterized, using Takesaki's result in the case of
von Neumann algebras. But a deeper analysis is needed since, at that stage the problem was not solved even in the
case of GW*-algebras. One of the reason is that the reduction of a GW*-algebra is not necessarily a GW*-algebra,
in contrast with the case of von Neumann algebras. For $\E_\NG$ to be a conditional expectations of $(\MM,\xi_0)$
w.r.t. $\NG$ (i.e., everywhwre defined on $\MM$), the following result given in \cite[Corollary 6.2]{inoue4}
holds.

\begin{lemma}\label{lemma_3.1}
Let $\MM$ be a closed O*-algebra on $\D$ in $\H$ such that $\MM_w'\D\subset\D$. Let $\xi_0$ be a strongly cyclic
and separating vector, in the sense that $\overline{\MM'_w\xi_0}=\H$. Suppose that $\NG$ is a closed
O*-subalgebra of $\MM$ satisfying
\begin{itemize}

\item[(i)] $\NG_w'\D\subset\D$;

\item[(ii)] $\NG\xi_0$ is essentially self-adjoint for $\NG$;

\item[(iii)] ${\Delta_{\xi_0}''}^{it}(\NG_w')'{\Delta_{\xi_0}''}^{-it}=(\NG_w')'$, $\forall t\in\mathbb{R}$, where ${\Delta_{\xi_0}''}$ is the modular operator of the left Hilbert algebra $(\MM_w')'\xi_0$;

\item[(iv)]  $\NG_{P_\NG}$ is a GW*-algebra on $P_\NG\D$, where    $P_\NG$ is a projection of $\H$ onto $\overline{\NG\xi_0}$.

\end{itemize}

Then $\E_{\NG}$ is a conditional expectation of $(\MM,\xi_0)$ w.r.t. $\NG$.

\end{lemma}

By this Lemma and Theorem \ref{thm_2.4} we deduce the following

\begin{theo}\label{thm_3.2}
Let $\MM$ be a closed O*-algebra on $\D$ in $\H$ with a strongly cyclic and separating vector $\xi_0$ such that
$\MM_w'\D\subset\D$ and let $\NG$ be an O*-subalgebra of $\MM$.

Suppose that $\NG$ is a GW*-algebra on a Fr\'echet domain $\D$ satisfying
\begin{itemize}

\item[(i)] $\NG\xi_0$ and $\left\langle\NG_w'P_\NG\D\right\rangle$ are essentially self-adjoint for $\NG$;

\item[(ii)] ${\Delta_{\xi_0}''}^{it}(\NG_w')'{\Delta_{\xi_0}''}^{-it}=(\NG_w')'$, $\forall t\in\mathbb{R}$.

\end{itemize}

Then $\E_\NG$ is a conditional expectation of $(\MM,\xi_0)$ w.r.t. $\NG$.

\end{theo}

By Lemma \ref{lemma_3.1} and Theorem \ref{thm_2.5} we deduce the following

\begin{theo}
Let $(\MM,\xi_0)$ be as in Theorem \ref{thm_3.2} and $\NG$ an O*-subalgebra of $\MM$. Suppose $\NG$ is a
GW*-algebra on $\D$ satisfying:
\begin{itemize}

\item[(i)] $t_\NG$ is defined by a sequence $\{T_n\}$ of essentially self-adjoint operators in $\NG$ whose
spectral projections leave the domain $\D$ invariant;

\item[(ii)] ${\Delta_{\xi_0}''}^{it}(\NG_w')'{\Delta_{\xi_0}''}^{-it}=(\NG_w')'$, $\forall t\in\mathbb{R}$.

\end{itemize}

Then $\E_\NG$ is a conditional expectation of $(\MM,\xi_0)$ w.r.t. $\NG$.

\end{theo}
\begin{proof} By Lemma \ref{lemma_2.5}, $\NG\xi_0$
and $\left\langle\NG_w'P_\NG\D\right\rangle$ are essentially self-adjoint for $\NG$; by Theorem \ref{thm_3.2} it
follows that $\E_\NG$ is a conditional expectation of $(\MM,\xi_0)$ w.r.t. $\NG$.

\end{proof}

\begin{cor} Let $(\MM,\xi_0)$ be as in Theorem \ref{thm_3.2} and $\NG$ an O*-subalgebra of $\MM$. Suppose $\NG$ is a
GW*-algebra on $\D^\infty(\overline{T})$ where $T$ is an essentially self-adjoint operator in $\NG$, and
${\Delta_{\xi_0}''}^{it}(\NG_w')'{\Delta_{\xi_0}''}^{-it}=(\NG_w')'$, $\forall t\in\mathbb{R}$. Then, $\E_\NG$ is
a conditional expectation of $(\MM,\xi_0)$ w.r.t. $\NG$.

\end{cor}

\begin{cor}
Let $(\MM,\xi_0)$ be as in Theorem \ref{thm_3.2} and $\NG$ an O*-subalgebra of $\MM$. Suppose $\NG$ is an
EW*-algebra on a Fr\'echet domain $\D$ satisfying
${\Delta_{\xi_0}''}^{it}(\NG_w')'{\Delta_{\xi_0}''}^{-it}=(\NG_w')'$, $\forall t\in\mathbb{R}$. Then $\E_\NG$ can
be extended to a conditional expectation $\E_{\bic{\NG}}$of $(\bic{\MM},\xi_0)$ w.r.t. $\bic{\NG}$.

\end{cor}

\bigskip
\noindent {\bf Acknowledgement} One of us (A.I.) acknowledges the hospitality of the Dipartimanto di Matematica
ed Applicazioni, Universit\`a di Palermo, where this work was performed.


\begin{thebibliography}{99}



\bibitem{dixmier} J. Dixmier, {\em Von Neumann algebras}, North-Holland Publ. Comp., Amsterdam, 1981.




\bibitem{inoue1} A. Inoue, {\em On a class of unbounded operator algebras},
Pacific J. Math. {\bf 65}, 11-95 (1976).



\bibitem{inoue3} A. Inoue, H. Ogi, M. Takakura, {\em Conditional expectations for unbounded operator algebras}, Int. J. Math. Math. Sci. 2007, Art. ID 80152, 22 pp.

\bibitem{inoue4} A. Inoue, H. Ogi, M. Takakura, {\em Conditional expectations for O*-algebras}, Contemporary Math., {\bf 427}, 225-234 (2007)


\bibitem{powers} R.T. Powers, {\em Self-adjoint algebras of unbounded operators}, Commun. Math. Phys., {\bf 21}, 85-124 (1971).

\bibitem{schmudgen} K. Schm\"{u}dgen,
{\em Unbounded Operator Algebras and Representation Theory}, Birkh\"{a}user-Verlag, Basel, 1990.


\bibitem{take} M. Takesaki, {\em Conditional expectations in a von Neumann algebra}, J. Funct. Anal., {\bf 9}, 306-327 (1972).





\end{thebibliography}
\end{document}